\documentclass[10pt, conference]{IEEEtran}
\usepackage{graphicx} % Required for inserting images
\usepackage{soul, xcolor}
\usepackage{comment}
\usepackage{balance}

\title{Post-Quantum Secure UE-to-UE Communications}

\author{Sanzida Hoque$^{\dagger}$, Abdullah Aydeger$^{\dagger}$ and Engin~Zeydan$^{\ast}$\\
$^{\dagger}$ Dept. of Electrical Engineering and Computer Science, Florida Institute of Technology, Melbourne, FL, USA. \\
$^{\ast}$Centre Tecnològic de Telecomunicacions de Catalunya (CTTC), Barcelona, Spain, 08860.\\
\protect Email: shoque2023@my.fit.edu, aaydeger@fit.edu, engin.zeydan@cttc.cat. 
}

\begin{document}

\maketitle

\begin{abstract}

%The rapid development of quantum computing poses a significant threat to the security of current cryptographic systems, including those used in User Equipment (UE) for mobile communications. Conventional cryptographic algorithms such as RSA and ECC are vulnerable to quantum computing attacks, which could jeopardize the confidentiality, integrity, and availability of sensitive data transmitted by UEs. This demo paper proposes the integration of Post-Quantum Cryptography (PQC) in UE to mitigate the risks of quantum attacks. We investigate the specific vulnerabilities of current UE security protocols and identify suitable PQC algorithms that can be integrated into existing UE architectures. We analyze the performance and resource overhead of these PQC algorithms from the perspectives of the UEs by running the virtual 5G testbed with multiple UEs. By addressing the challenges and opportunities of using PQC in UE, this research aims to pave the way for the development of quantum-resistant mobile devices and secure the future of wireless communications.

The rapid development of quantum computing poses a significant threat to the security of current cryptographic systems, including those used in User Equipment (UE) for mobile communications. Conventional cryptographic algorithms such as Rivest–Shamir–Adleman (RSA) and Elliptic curve cryptography (ECC) are vulnerable to quantum computing attacks, which could jeopardize the confidentiality, integrity, and availability of sensitive data transmitted by UEs. This demo paper proposes the integration of Post-Quantum Cryptography (PQC) in TLS for UE Communication to mitigate the risks of quantum attacks. We present our setup and explain each of the components used. We also provide the entire workflow of the demo for other researchers to replicate the same setup. By addressing the implementation of PQC within a 5G network to secure UE-to-UE communication, this research aims to pave the way for developing quantum-resistant mobile devices and securing the future of wireless communications. %By addressing the post-quantum threats using PQC in UE, this research aims to pave the way for the development of quantum-resistant mobile devices and secure the future of wireless communications.
% We also provide the entire workflow of the demo for other researchers to replicate the same setup.

\vspace{-0.2in}
\end{abstract}

\section{Introduction}

Quantum computing poses a significant threat to the security of modern cryptographic systems, including those used in mobile communications. User Equipment (UE) such as smartphones and other wireless devices rely heavily on cryptographic protocols for secure communication, authentication, and data protection \cite{hoque2024exploring}. However, the cryptographic algorithms commonly used in UE, such as Rivest–Shamir–Adleman (RSA) and Elliptic Curve Cryptography (ECC), are vulnerable to attacks by sufficiently powerful quantum computers. These attacks, which exploit algorithms such as the Shor algorithm \cite{shor1994algorithms}, could break the underlying mathematical problems on which these cryptographic methods are based, rendering them ineffective. 

Therefore, the advent of quantum computing necessitates the development and integration of Post-Quantum Cryptography (PQC) in UE. PQC comprises a set of cryptographic algorithms that are believed to be resistant to attacks from both classical and quantum computers \cite{aydegertowards}. These algorithms leverage mathematical problems that are considered difficult for even quantum computers to solve, providing a more robust foundation for security in the quantum era \cite{hanna2023performance}. The integration of PQC in UE poses a particular challenge due to the limited resources of mobile devices, such as limited computing power, memory, and battery life. Another hurdle is ensuring compatibility with existing network infrastructure and protocols. However, the potential benefits of PQC in safeguarding user data and privacy in the face of quantum threats make it an important area of research.

This paper leverages the recent advances in PQC to enable secure and scalable interactions between UE in a 5G testbed. The testbed demonstrates the practical application of these technologies by showcasing real-time post-quantum secure communication among UE devices. Specifically, it uses Kyber/ML Key Encapsulation Mechanism (KEM) \cite{Pathum_2024}, a NIST-standardized KEM, for key management over TLS. By integrating the Kyber with wolfSSL, we illustrate the process for establishing secure communication channels resistant to quantum attacks between UEs in a virtual 5G setup developed using Open5gs and UERANSIM.
%The testbed exemplifies the practical use of these technologies, demonstrating real-time post-quantum secure communication among UE devices, specifically using Kyber/ML KEM \cite{Pathum_2024}, NIST standardized Key Encapsulation Mechanism (KEM), for key management and TLS encryption. By integrating the Kyber algorithm with wolfSSL over TLS, we showcase the process for establishing secure communication channels that are resistant to quantum attacks between UEs in a virtual 5G setup, developed using Open5gs and UERANSIM. 
 The research aim is to establish the groundwork for the development of quantum-resilient mobile devices that ensure the security and privacy of user data in the upcoming era of quantum computing.  %It is important to note that, pqc integrated wolfssl do allow integration of post-quantum signature algorithm for authentication, like Dilithium or Falcon. However, this work do not include these certificates and focus on quantum safe key exchange.  %Among the main cryptographic steps that includes quantum safe encryption, quantum safe quthentical and quantum safe key exchange, our focus is the last one. It is important to note that integration of pqc digital signature algorithm is possible in pqc integrated wolfssl. This set up use commonly used encryption like AES 128 and digital signature like ecc.

%This paper addresses the challenges and opportunities of integrating PQC into UE. First, we examine the specific vulnerabilities of current UE security protocols against quantum attacks. Then, we identify and evaluate suitable PQC algorithms that can be efficiently implemented on resource-constrained devices. We analyze the trade-offs between security, performance, and resource consumption for different PQC algorithms. Finally, we outline a framework for integrating PQC into UE architectures, taking into account factors such as protocol compatibility, key management, and performance optimization. By addressing these challenges and proposing concrete solutions, this research aims to pave the way for the development of quantum-resilient mobile devices that ensure the security and privacy of user data in the upcoming era of quantum computing.

%\section{Post-Quantum Cryptography for UE Communications}

\footnotetext{Accepted for NoF 2024 Demo track.}

\section{5G based Testbed for PQC Analysis}
%== update as it fits == rest taken from the doc file you shared earlier==

%We use the latest technologies to emulate PQC-secured UE-to-UE communication. 
In order to display our approach in this paper, we use Dell Desktops equipped with an i7 CPU with 20 cores, 32 GB RAM, and 512 GB SSD storage. We deploy different components of this setup in separate virtual machines (VMs) on the KVM hypervisor. The operating system used in each VM is Ubuntu Server version 20.04.6 LTE. Each VM has 4 GB of memory and two CPU cores. We use Open Vswitch (OVS)\footnote{Online: https://github.com/openvswitch/ovs, Available: July 2024} to set up and manage the virtual network interfaces and bridges required to simulate the connections between different network components. 
% We chose this operating system because it is open source, has networking capabilities, and is compatible with a variety of security solutions.

\begin{figure}[h!]
\centering
\includegraphics[width=\linewidth]{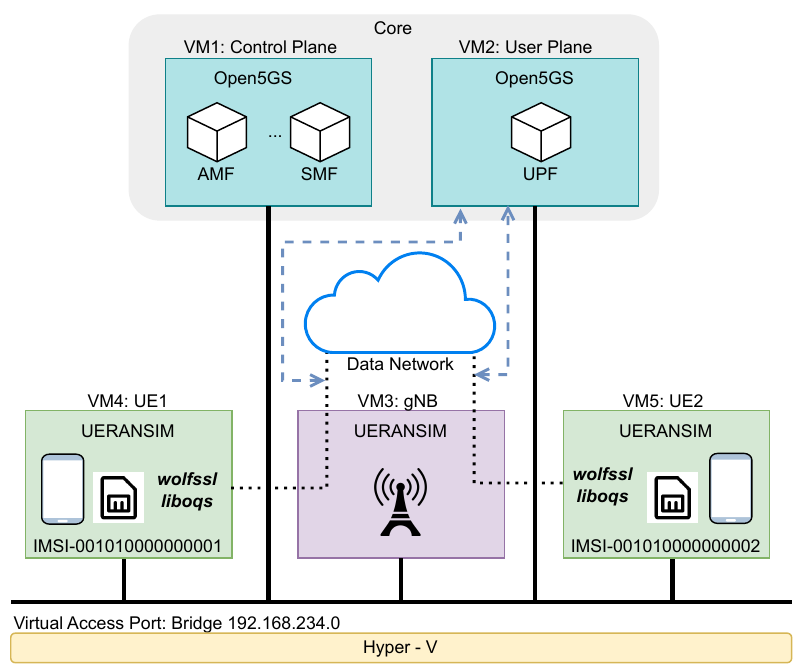}
\vspace{-0.1in}
\caption{5G simulation environment used within the demo}
\vspace{-0.2in}
\label{fig:arch}
\end{figure}
\subsection{5G Testbed}

For the 5G setup, we use open-source technologies such as UERANSIM and Open5Gs. UERANSIM \footnote{Online: https://github.com/aligungr/UERANSIM, Available: July 2024} is the simulator for 5G UE and gNodeB (gNB), while Open5GS \footnote{Online: https://github.com/open5gs/open5gs, Available: July 2024} is a project that provides a complete 5G Core (5GC) network implementation and enables testing and development of 5G networks. The components are deployed in various VMs. The most critical components of the 5G-based testbed are described below:

    \textbf{UE}: These are devices such as smartphones, tablets, IoT devices, etc., that are connected to the network. UERANSIM is used to create a realistic simulation environment. We use two UEs, one of which acts as a server and the other as a client. However, these two UEs can also work in reverse. UERANSIM provides a TUN interface to utilize the UE's internet connectivity. 
    
    \textbf{gNB}: The Next Generation Node B, gNB for short, manages the radio access portion of the network and provides an interface that connects UEs to the 5G core network. UERANSIM is also used here to simulate the 5G gNB and test the connectivity and performance of 5G communication with UEs.
    %is the 5G version of the evolved Node B, eNB for short, which was used in 4G LTE networks. It
    
    \textbf{Core Network}: The 5G Core (5GC) is responsible for various tasks within the mobile network and enables communication. The 5G Core network is cloud-native and fully software-based, enabling greater agility and flexibility when deploying on different cloud infrastructures. Two main planes characterize the architecture: the User Plane (UP) and the Control Plane (CP). The CP is responsible for managing the signaling and network functions required to establish and maintain sessions, manage mobility, and enforce policies, while the UP manages the actual traffic to ensure the transmission of user data. Key components of CP are (i) AMF (Access and Mobility Management Function) authenticates the UE and establishes a connection. When a user moves between different cells or networks, AMF handles the handover process to ensure continuous connectivity. (ii) SMF (Session Management Function) establishes a data session and sets up an interface with the UPF (User Plane Function), which is the central component of UP and serves as the gateway to data network (DN) for user traffic. This function is responsible for packet forwarding, routing, and traffic shaping. UP also includes the DN, which consists of application servers, Internet access, and other services through which the UE communicates. CP includes other functions such as PCF (Policy Control Function), UDM (Unified Data Management), and UDR (Unified Data Repository).

    In this setup, UE, gNB, CP, and UP have been deployed in four different VMs. The setup is shown in the Figure \ref{fig:arch}. First, each UE connects to the network using the AMF, which sets up authentication and authorization. A PDU(Packet Data Unit) session is then established, with the SMF configuring the UPF for data processing. During data transmission, a data unit, known as a data packet, is transmitted from UE1 to the gNB. The gNB then transmits the data packet to the UPF via the GTP-U protocol. The UPF forwards the packet via the DN to the other UPF, which serves the gNB of UE2. The gNB at the destination then transmits the packet to UE2. If both UEs are in the same local network, it may not be necessary to include a DN. Instead, the UPF can manage the local routing.

%Data transfer between User Equipments (UEs) in a 5G network entails multiple technical procedures. At first, every User Equipment (UE) connects to the network by means of the Access and Mobility Management Function (AMF), which sets up authentication and authorization. Subsequently, a Packet Data Unit (PDU) session is established, whereby the Session Management Function (SMF) configures the User Plane Function (UPF) to handle data. The control plane is responsible for handling signaling processes related to the distribution of resources and the establishment of connections. During the process of transmitting data, a unit of data called a data packet is delivered from UE1 to the gNodeB (gNB). The gNB then transmits the data packet to the UPF via the GTP-U protocol. The User Plane Function (UPF) directs the packet over the data network (DN) to the UPF that serves UE2's gNB. The destination gNB subsequently transfers the packet to UE2. The control plane is responsible for maintaining Quality of Service (QoS) and managing handovers when User Equipment (UEs) move between cells. It enables uninterrupted connectivity and efficient data flow throughout the 5G infrastructure.

\subsection{PQC Secured Communication}
% OQS is a member of the PQC Alliance, affiliated with the Linux Foundation.
The Open Quantum Safe (OQS) project is a publicly available initiative to facilitate the transition to cryptographic methods, resistant to quantum computing \cite{stebila2016post}. OQS provided liboqs \footnote{Online: https://github.com/open-quantum-safe/liboqs, Available: July 2024}, a publicly available C library that provides cryptographic algorithms that are resistant to quantum attacks. wolfSSL\footnote{Online: https://github.com/wolfSSL/wolfssl, Available: July 2024} is a light version of OpenSSL, an alternative SSL/TLS library. The wolfSSL team integrates experimental PQC algorithms, such as Kyber, into the wolfSSL library. In this experiment, the wolfSSL library is used to establish TLS connections, and the liboqs library is used to generate Kyber keys in both UEs, as indicated in Figure \ref{fig:arch}. It can be noted that wolfSSL enables the use of post-quantum signature algorithms, such as Dilithium or Falcon, for authentication. However, this work is specifically centered around quantum-safe key exchange.

\begin{figure}[h!]
\centering
\includegraphics[width=\linewidth]{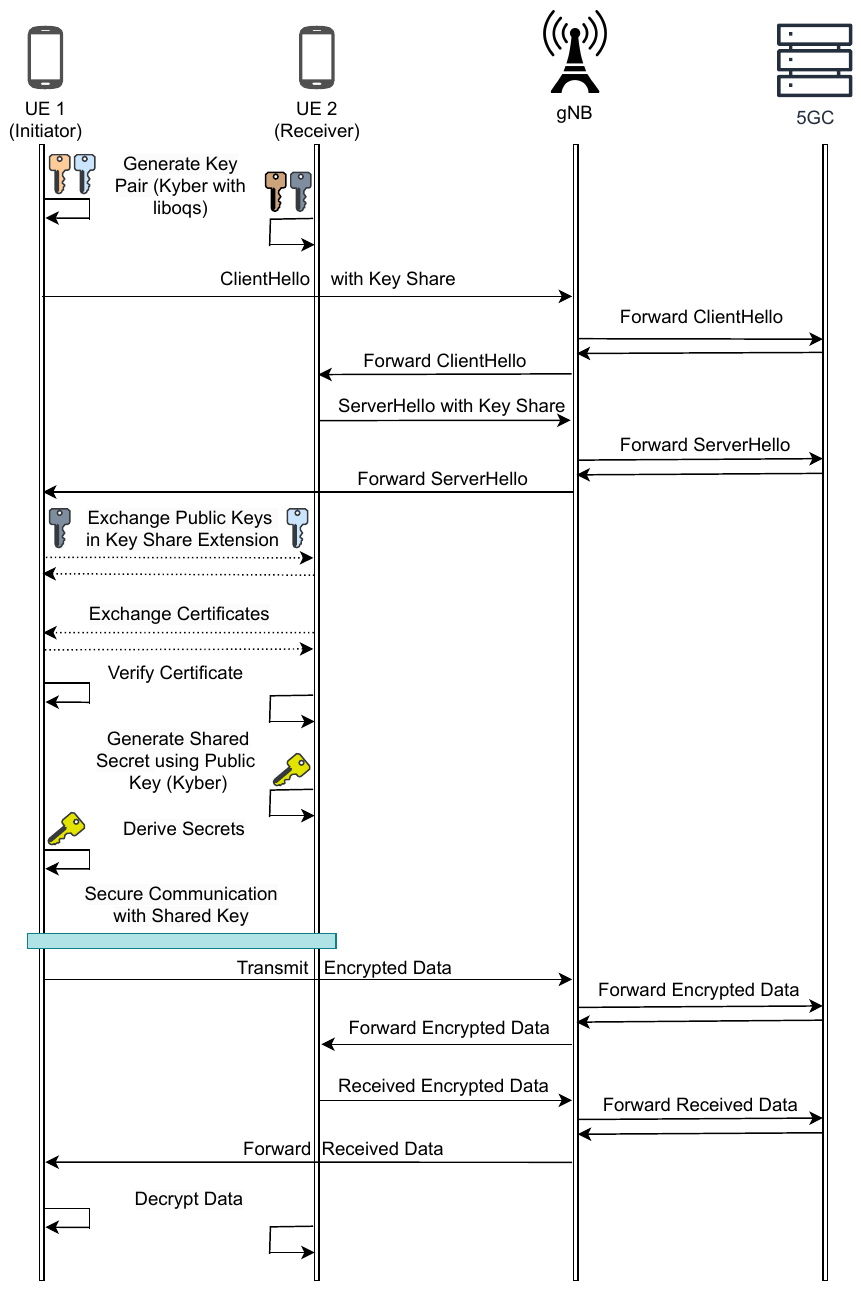}
%\vspace{-0.2in}
\caption{Kyber/ML KEM enabled UE to UE Communication}
\vspace{-0.2in}
\label{fig:6gd}
\end{figure}

\section{Demo Workflow }

The sequence diagram of the quantum-safe UE-to-UE communication experimented in this paper is shown in Figure \ref{fig:6gd}. The most important steps in this demo are described below: %\cite{Merricx, Pathum_2024}:

\textbf{(i) UE Registration}: First, the UEs are registered on the 5GC via the WebUI application. Then we run the configuration files of the UEs in UERANSIM, and both UEs (UE1 and UE2) send registration requests to the gNB, which serves as an intermediary between the UEs and the core network and is shown as VM3 in Figure \ref{fig:arch}. The gNB transmits these registration requests to the core network (VM1), where authentication and authorization take place. The core network validates the credentials of each UE and sends confirmation messages to the gNB after successful authentication. The gNB then sends these confirmation signals to UE1 (VM4) and UE2 (VM5), thereby completing the registration process and establishing a secure connection to the network. It is important to note that the scope of this study does not include the integration of Kyber during UE registration, rather it focuses on its integration during TLS establishment for the key exchange which begins with the following steps.

\textbf{(ii) Key Pair Generation}: To generate the Kyber key pairs, both UEs use the liboqs library after completing the registration process. The generation of key pairs is crucial for ensuring secure communication, as these keys are used for secure key exchange during the TLS handshake, encapsulating and decapsulating symmetric keys that are then used for session encryption. UE1 and UE2 generate their public and private keys autonomously by executing the Kyber key generation function provided by liboqs, as shown in the first steps in Figure \ref{fig:6gd}.

\textbf{(iii) TLS Handshake Initialization}: To further secure the communication, UE1 initiates a TLS 1.3 handshake by sending a Client Hello message to UE2. After receiving the Client Hello message, UE2 responds with a Server Hello message. These messages exchange critical cryptographic information such as public key/ciphertext, certificates, etc., as shown in Figure \ref{fig:6gd} with a dotted line to indicate that this information is contained in the ClientHello and ServerHello packets and has already been exchanged \cite{Merricx, Westerbaan_2023}. %containing the Key Share extension and other cryptographic parameters. This message contains the necessary cryptographic information and preferences, such as public key/ciphertext, certificates, etc., shown with a dotted line in Figure\ref{fig:6gd} to indicate that this information is contained in the ClientHello and ServerHello packets and has already been exchanged. %According to draft-ietf-tls-hybrid-design-10 \cite{Merricx, Westerbaan_2023}, the extensions data fields in the ClientHello and ServerHello messages exchanged during TLS handshake initialization contain the public key/ciphertext as KeyShareEntry, which is required during secret derivation stage. The KeyShareEntry contains the key exchange field, which is formed by combining the key exchange fields of the individual algorithms. 
This handshake ensures that both UEs agree on the cryptographic parameters and establish initial security parameters that are proposed in RFC 8446 \footnote{Online: https://www.rfc-editor.org/rfc/rfc5280, Available: May 2008}
%\cite{Cooper_Santesson_Farrell_Boeyen_Housley_Polk_2008}.
%According to draft-ietf-tls-hybrid-design-09[3], public key or ciphertext is represented as KeyShareEntry in the extensions_data fields. This KeyShareEntry will then contain key_exchange field which is the concatenation of the key_exchange field for each of the constituent algorithms.
\textit{\textbf{(a) Public Key Exchange}}: In this experiment, Kyber is used in the key exchange phase of TLS. To create a secure communication channel, UE1 and UE2 share their public keys over the 5G network using the secure channels established during the registration process. In this testbed, UE1 transmits its public key to UE2, and in return, UE2 sends its public key to UE1. This key sharing allows each UE to use the other party's public key for subsequent cryptographic procedures, as described in RFC 8446   \footnote{Online: https://datatracker.ietf.org/doc/rfc8446/, Available: Mar 2020}.
%\cite{rfc8446}.
\textit{\textbf{(b) Certificate Exchange}}: During this stage, UE1 and UE2 exchange digital certificates to mutually verify and validate their respective identities. UE1 sends its certificate to UE2 and UE2 sends it to UE1 (RFC 5280) \cite{housley2018internationalization}.%\footnote{Online: https://datatracker.ietf.org/doc/html/rfc5280, Available: May 2008}%\cite{housley2018internationalization}.
%Certificates, commonly given by a trusted Certificate Authority (CA), store public keys and are utilized to authenticate the identification of the communicating parties, guaranteeing that both UEs are genuinely the individuals they assert to be (RFC 5280) \cite{housley2018internationalization}%\footnote{Online: https://datatracker.ietf.org/doc/html/rfc5280, Available: May 2008}%\cite{housley2018internationalization}. % It can be noted that PQC-integrated wolfSSL allows for the use of post-quantum signature algorithms, such as Dilithium or Falcon, for authentication. However, this work is specifically centered around quantum-safe key exchange. %It is important to note that, pqc integrated wolfssl do allow integration of post-quantum signature algorithm for authentication, like Dilithium or Falcon. However, this work do not include these certificates and focus on quantum safe key exchange.  %Among the main cryptographic steps that includes quantum safe encryption, quantum safe quthentical and quantum safe key exchange, our focus is the last one. It is important to note that integration of pqc digital signature algorithm is possible in pqc integrated wolfssl. This set up use commonly used encryption like AES 128 and digital signature like ecc.

\textbf{(iv) Shared Secret Generation}: Shared secrets are generated by both UEs by utilizing the Kyber with the exchanged public keys and their respective private keys. UE1 utilizes the public key of UE2 and its own private key to create a shared secret and a corresponding ciphertext. UE2 utilizes the received ciphertext and its private key to obtain the identical shared secret. The shared secret is an essential element in the establishment of a secure communication channel using the Kyber KEM. The library used in the experiment, liboqs, supports Kyber/ML-KEM in both hybrid and conventional modes, as well as varying security levels.

\textbf{(v) Shared Secret Derivation}: The shared secrets are derived by both UEs based on the information exchanged during the TLS handshake. The derivation process results in shared keys that are used for the encryption and decryption of the subsequent communication. %UE 1 and UE 2 use these keys to establish a secure communication channel for the transmission of data, as described in RFC 8446 \cite{rfc8446}.

\textbf{(vi) Handshake Completion}: After the successful derivation of the session secrets, the handshake is completed, and a Kyber-enabled secure communication channel is established, as presented with a sky blue line in Figure \ref{fig:6gd}. 

\textbf{(vii) Secure Message Exchange}: Now that the secure communication channel has been set up, UE1 and UE2 can exchange encrypted messages. Each message is encrypted and digitally signed before sending, decrypted, and validated after receipt. Using Kyber PQC guarantees that security and privacy are maintained throughout the communication process, effectively safeguarding against the risks posed by future advances in quantum computing.

\balance

\bibliographystyle{IEEEtran}
\bibliography{references}

\begin{thebibliography}{1}
\providecommand{\url}[1]{#1}
\csname url@samestyle\endcsname
\providecommand{\newblock}{\relax}
\providecommand{\bibinfo}[2]{#2}
\providecommand{\BIBentrySTDinterwordspacing}{\spaceskip=0pt\relax}
\providecommand{\BIBentryALTinterwordstretchfactor}{4}
\providecommand{\BIBentryALTinterwordspacing}{\spaceskip=\fontdimen2\font plus
\BIBentryALTinterwordstretchfactor\fontdimen3\font minus \fontdimen4\font\relax}
\providecommand{\BIBforeignlanguage}[2]{{%
\expandafter\ifx\csname l@#1\endcsname\relax
\typeout{** WARNING: IEEEtran.bst: No hyphenation pattern has been}%
\typeout{** loaded for the language `#1'. Using the pattern for}%
\typeout{** the default language instead.}%
\else
\language=\csname l@#1\endcsname
\fi
#2}}
\providecommand{\BIBdecl}{\relax}
\BIBdecl

\bibitem{hoque2024exploring}
S.~Hoque, A.~Aydeger, and E.~Zeydan, ``{Exploring Post Quantum Cryptography with Quantum Key Distribution for Sustainable Mobile Network Architecture Design},'' \emph{arXiv preprint arXiv:2404.10602}, 2024.

\bibitem{shor1994algorithms}
P.~W. Shor, ``Algorithms for quantum computation: discrete logarithms and factoring,'' in \emph{Proceedings 35th annual symposium on foundations of computer science}.\hskip 1em plus 0.5em minus 0.4em\relax Ieee, 1994, pp. 124--134.

\bibitem{aydegertowards}
A.~Aydeger, E.~Zeydan, A.~K. Yadav, K.~T. Hemachandra, and M.~Liyanage, ``{Towards a Quantum-Resilient Future: Strategies for Transitioning to Post-Quantum Cryptography},'' in \emph{2024 IEEE Network of Future (NoF)}.\hskip 1em plus 0.5em minus 0.4em\relax IEEE, 2024.

\bibitem{hanna2023performance}
Y.~Hanna, D.~Pineda, K.~Akkaya, A.~Aydeger, R.~Harrilal-Parchment, and H.~Albalawi, ``Performance evaluation of secure and privacy-preserving dns at the 5g edge,'' in \emph{2023 IEEE 20th International Conference on Mobile Ad Hoc and Smart Systems (MASS)}.\hskip 1em plus 0.5em minus 0.4em\relax IEEE, 2023, pp. 89--97.

\bibitem{Pathum_2024}
\BIBentryALTinterwordspacing
U.~Pathum, ``{Crystals Kyber: The key to Post-Quantum Encryption},'' Jun 2024. [Online]. Available: \url{https://medium.com/@hwupathum/crystals-kyber-the-key-to-post-quantum-encryption-3154b305e7bd}
\BIBentrySTDinterwordspacing

\bibitem{stebila2016post}
D.~Stebila and M.~Mosca, ``Post-quantum key exchange for the internet and the open quantum safe project,'' in \emph{International Conference on Selected Areas in Cryptography}.\hskip 1em plus 0.5em minus 0.4em\relax Springer, 2016, pp. 14--37.

\bibitem{Merricx}
\BIBentryALTinterwordspacing
``{RWPQC 2024 CTF},'' March 2024. [Online]. Available: \url{https://merri.cx/rwpqc2024-ctf/}
\BIBentrySTDinterwordspacing

\bibitem{Westerbaan_2023}
\BIBentryALTinterwordspacing
B.~E. Westerbaan, ``\BIBforeignlanguage{en}{{X25519Kyber768Draft00 hybrid post-quantum key agreement}},'' Sep. 2023. [Online]. Available: \url{https://datatracker.ietf.org/doc/html/draft-tls-westerbaan-xyber768d00-03}
\BIBentrySTDinterwordspacing

\bibitem{housley2018internationalization}
R.~Housley, ``{Internationalization updates to RFC 5280},'' 2018.

\end{thebibliography}

\end{document}